\documentclass[aps,prl,preprint,superscriptaddress,showpacs,preprintnumbers,amsmath,amssymb]{revtex4}
\usepackage{graphicx}
\usepackage{amsmath}
\usepackage{longtable}
\usepackage{bm}

\begin{document}

\title{Spin-Electron-Phonon Excitation in Re-based Half-Metallic Double Perovskites}

\author{A. F. Garc\'{i}a-Flores}
\affiliation{Instituto de F\'{i}sica "Gleb Wataghin", UNICAMP, 13083-970 Campinas, S\~{a}o Paulo, Brazil}

\author{A. F. L. Moreira}
\affiliation{Instituto de F\'{i}sica "Gleb Wataghin", UNICAMP, 13083-970 Campinas, S\~{a}o Paulo, Brazil}

\author{U. F. Kaneko}
\affiliation{Instituto de F\'{i}sica "Gleb Wataghin", UNICAMP, 13083-970 Campinas, S\~{a}o Paulo, Brazil}

\author{F. M. Ardito}
\affiliation{Instituto de F\'{i}sica "Gleb Wataghin", UNICAMP, 13083-970 Campinas, S\~{a}o Paulo, Brazil}

\author{H. Terashita}
\affiliation{Instituto de F\'{i}sica "Gleb Wataghin", UNICAMP, 13083-970 Campinas, S\~{a}o Paulo, Brazil}

\author{M. T. D. Orlando}
\affiliation{Departamento de F\'{i}sica e Qu\'{i}mica, UFES, 29075-910 Vit\'{o}ria, Esp\'{i}rito Santo, Brazil}

\author{J. Gopalakrishnan}
\affiliation{Solid State and Structural Chemistry Unit, Indian Institute of Science, 560012 Bangalore, India}

\author{K. Ramesha}
\affiliation{Solid State and Structural Chemistry Unit, Indian Institute of Science, 560012 Bangalore, India}

\author{E. Granado}
\affiliation{Instituto de F\'{i}sica "Gleb Wataghin", UNICAMP, 13083-970 Campinas, S\~{a}o Paulo, Brazil}

\begin{abstract}

A remarkable hardening ($\sim 30$ cm$^{-1}$) of the normal mode of vibration associated with the symmetric stretching of the oxygen octahedra for the Ba$_2$FeReO$_6$ and Sr$_2$CrReO$_6$ double perovskites is observed below the corresponding magnetic ordering temperatures. The very large magnitude of this effect and its absence for the anti-symmetric stretching mode provide evidence against a conventional spin-phonon coupling mechanism. Our observations are consistent with a collective excitation formed by the combination of the vibrational mode with oscillations of local $3d$ and $5d$ occupations and spin magnitudes.

\end{abstract}


\maketitle{}

Double perovskites are stable and simple crystal structures that provide a convenient network to investigate fundamental magnetic interactions in solids. Some compounds, such as $A_2$FeMoO$_6$ and $A_2 M$ReO$_6$ ($A=$ Sr, Ba, $M=$ Fe, Cr), have been widely studied due to its spin-polarized conduction electrons arising from a half-metallic ferrimagnetic ground states with ordering transitions at $T_c$ above room temperature, making them potential materials for spintronics devices \cite{Kobayashi1,review}. In these materials the minority-spin conduction electrons have a mixed $M$ $3d$ and Mo $4d$ or Re $5d$ character and mediate a double-exchange-like ferromagnetic interaction between the localized $3d$ majority spins. For the Re-based compounds, an additional ingredient that further enriches the physics is the partly localized character of the $5d$ moments. In fact, the high $T_c \sim 520$ K for the non-metallic Ca$_2$FeReO$_6$ \cite{GranadoCFRO,Kato1} indicates that alternative exchange mechanisms not involving conduction electrons are dominant, and the large coercive fields $H_c$ of several tesla as opposed to a few Oersted for Sr$_2$FeMoO$_6$ have been regarded a clear manifestation of the large spin-orbit coupling of the Re $5d$ electrons \cite{review}. Finally, these Re based double perovskites were reported to be at the frontier of a metal-insulator transition, driven by spin-orbit coupling and unexpectedly large Re $5d$ correlations \cite{review,GranadoCFRO,Kato1,Iwasawa,Sikora,Azimonte}. 

The hybridization of the $M$ $3d$, O $2p$, and Re $5d$ minority-spin levels at the conduction band captured by band structure calculations \cite{Kobayashi,Wu,Jeng,Vaitheeswaran} is related to the $M$-O and Re-O distances, therefore the oxygen positions may be important parameters to the magnetism of this family \cite{Solovyev}. Particularly, the population of $M$ $3d$ and Re $5d$ minority-spin levels below the Fermi level are expected to be strongly dependent on the oxygen positions. The transition-metal valence states and atomic spins should be also sensitive to the {\it dynamical} oxygen displacements brought by phonons in the lattice. This would lead to collective excitations formed by the combination of vibrational modes and oscillations of local $3d$ and $5d$ occupations and spin magnitudes. The parent $Fm \bar{3} m$ cubic double perovskite structure is particularly simple and ideally suited for the spectroscopic observation of such spin-electron-phonon collective excitation. This structure allows for four Raman-active vibration modes ($A_{1g}+E_{g}+2F_{2g}$), three of them ($A_{1g}+E_{g}+F_{2g}$) involving mostly displacements of the oxygen ions that compose the octahedral environment of the transition-metal ions (see Fig. 1(a)) \cite{Iliev}, while the remaining lower-frequency $F_{2g}$ mode involves mostly displacements of the $A$ ions. The $A_{1g}$ symmetric stretching mode is of particular interest since it modulates the volume of the $M$O$_6$ and ReO$_6$ octahedra and therefore the chemical pressure of the oxygen cages onto the transition-metal ions. In this work, we investigate the phonon behavior of Re-based half-metallic ferrimagnetic double perovskites through the magnetic ordering transition by means of Raman spectroscopy. Ba$_2$FeReO$_6$ (BFRO) and Sr$_2$CrReO$_6$ (SCRO) were chosen for being ferrimagnets with $T_c=305$ \cite{Azimonte,Prellier,Teresa1} and $\sim 600-635$ K \cite{Kato,Teresa3,Blasco}, respectively, with cubic double perovskite structures at the paramagnetic state \cite{Gopalakrishnan,Rammeh,Azimonte,Teresa3,Blasco}. We report a record-high energy renormalization of the symmetric stretching mode of the oxygen octahedra below $T_c$ for both compounds, providing strong experimental evidence of the formation of the spin-electron-phonon collective excitation.

\begin{figure}
\includegraphics[width=1 \textwidth]{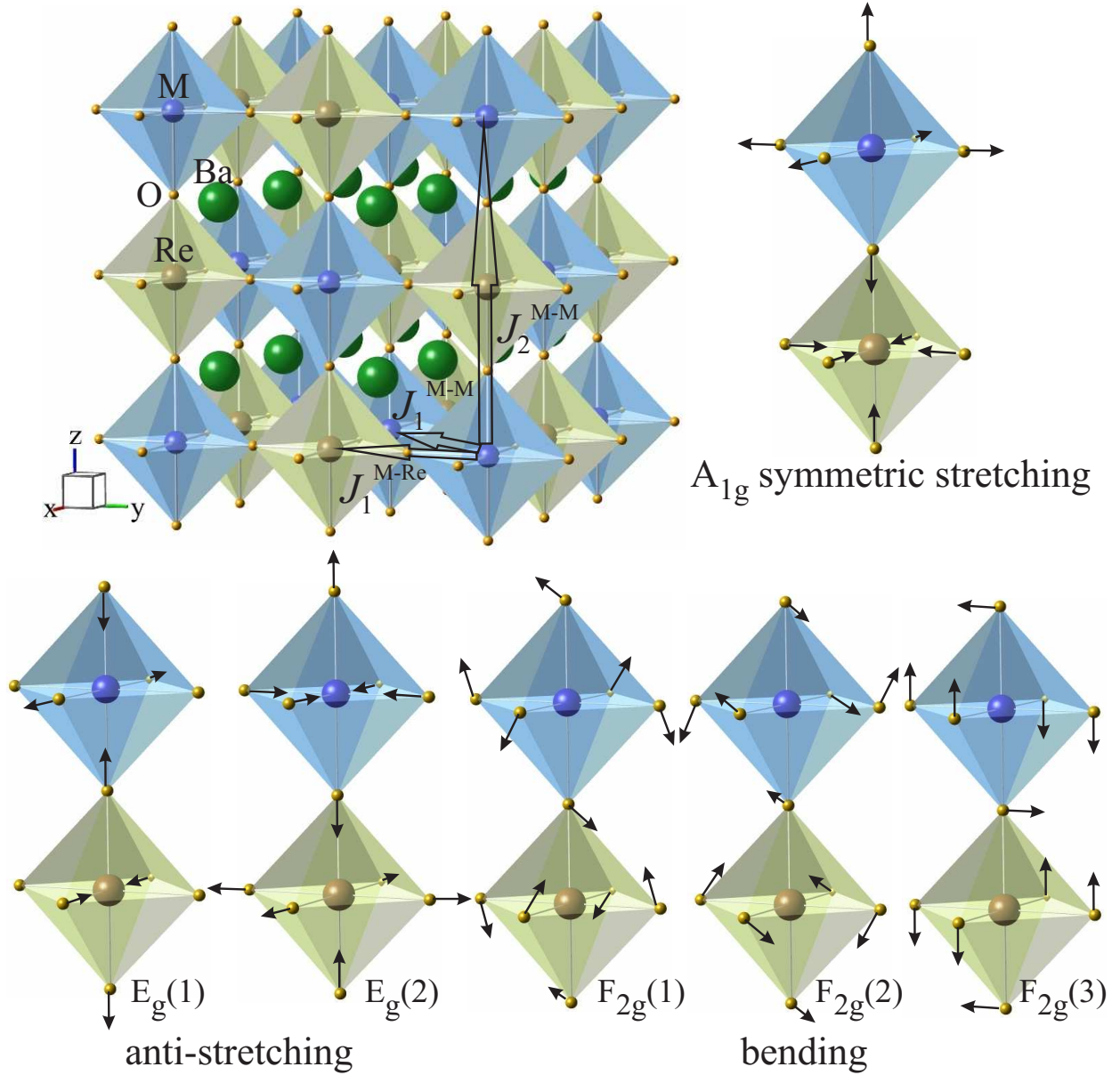}
\caption{\label{structure} (Color online) Crystal structure and relevant Raman-active modes in $Fm\bar{3}m$ double perovskites. The $J_1^{M-M}$, $J_2^{M-M}$ and $J_1^{M-Re}$ ($M=$ Fe, Cr) exchange interactions are also indicated (see text).}
\end{figure}

The synthesis and characterization of the BFRO ceramic sample are described elsewhere \cite{Prellier,Gopalakrishnan,Azimonte}. SCRO was synthesized by solid state reaction. A mixture of SrO, Cr$_2$O$_3$, ReO$_2$ and ReO$_3$ powders in  $2:0.5:0.4:0.6$ ratio was ground and pelletized in Ar atmosphere. This mixture was wrapped in gold foil and sealed in an evacuated quartz tube. The sample was sintered for a total time of 154 h at 1030 $^{\circ}$C with two intermediate grinding steps. High resolution synchrotron x-ray powder diffractions experiments in SCRO were performed in the XPD beamline of the Brazilian Synchrotron Laboratory (LNLS) \cite{Ferreira} with $\lambda=1.2390$ \AA\ and indicate a single phase with cubic double perovskite structure (space group $Fm\bar{3}m$) at 350 K. Magnetization measurements were performed with a commercial vibrating sample magnetometer under field cooling and confirmed the magnetic ordering below $T_c=605$ K for SCRO. The degrees of antisite disorder are 4(1) \% for BFRO \cite{Azimonte} and 16(1) \% for SCRO. The relatively high degree of antisite disorder for SCRO is similar to previous reports \cite{Kato,Teresa3,Blasco}. Raman scattering measurements were performed using a Jobin Yvon T64000 triple grating spectrometer equipped with a $L$N$_{2}$-cooled multichannel CCD detector. The Raman spectra were excited with the 488 nm Ar$^{+}$ laser line in a quasi-backscattering configuration. The incident laser power was kept below 10 mW focused into a spot of $\sim 100$ $\mu$m of diameter to avoid local heating. The $T$-dependent measurements were carried out by mounting the samples with fresh broken surfaces on a cold finger of a closed-cycle He refrigerator. The  magneto-Raman and magnetoresistance measurements in BFRO were performed using the same commercial superconducting optical magnetocryostat and the four-probe method.

\begin{figure}
\includegraphics[width=0.8 \textwidth]{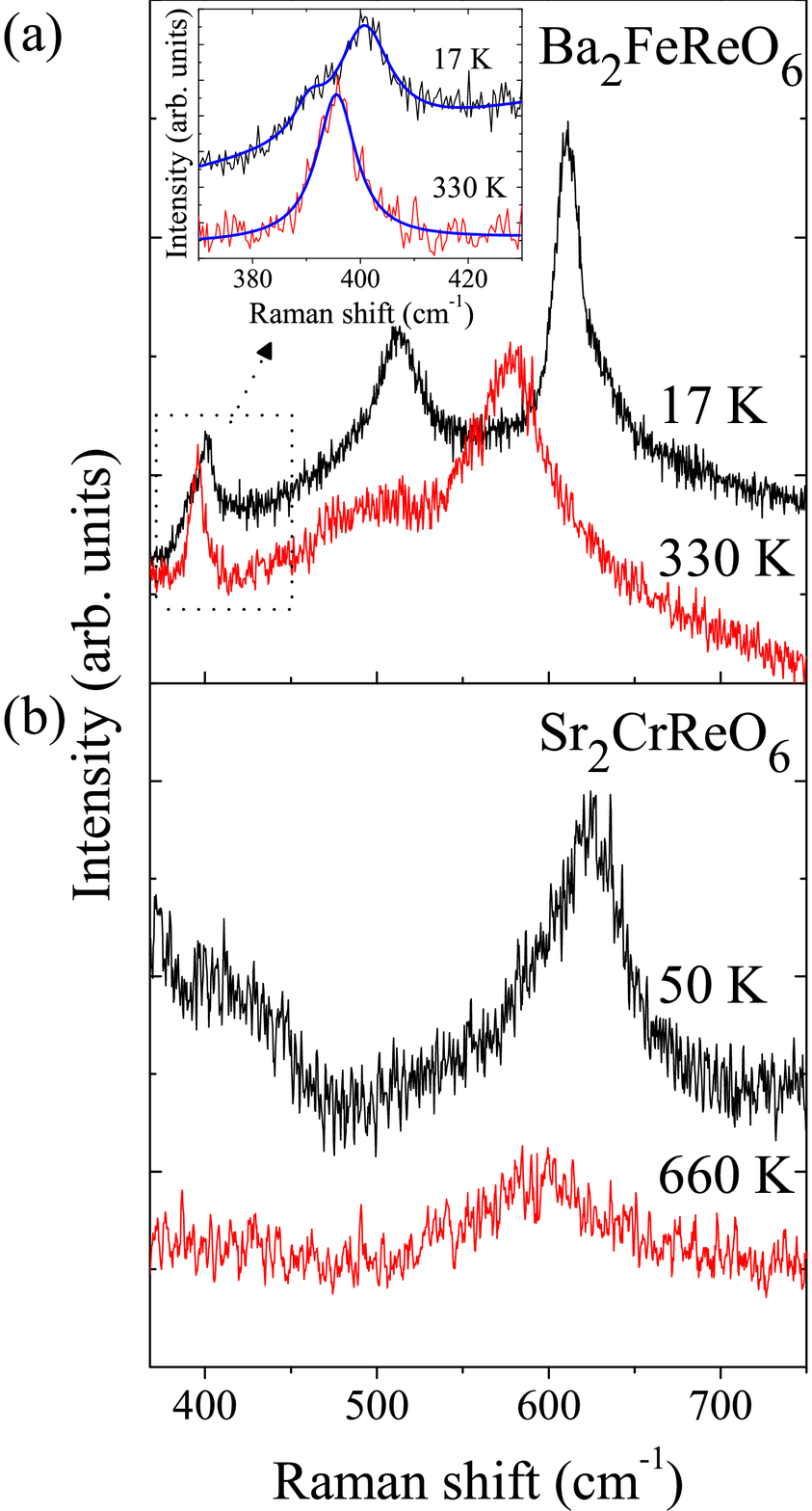}
\caption{\label{Spectra} (Color online) Raman spectra of polycrystalline Ba$_{2}$FeReO$_{6}$ ($T_c=305$ K) at 17 and 330 K (a) and Sr$_{2}$CrReO$_{6}$ ($T_c \sim 605$ K) at 30 and 630 K (b). The inset shows in detail the region of the $F_{2g}$ bending mode of Ba$_{2}$FeReO$_{6}$ at 330 and 17 K. Smooth lines are fits to this region using one and two Lorentzian peaks, respectively.}
\end{figure}

Figure \ref{Spectra}(a) displays the Raman spectra of BFRO at 17 and 330 K. Clear peaks at 395, 495, and 580 cm$^{-1}$ were observed at 330 K. Polarization analysis (not shown) confirms they are associated with the $F_{2g}$ bending, $E_g$ anti-stretching and $A_{1g}$ symmetric stretching modes, respectively (see Fig. \ref{structure}).
Figure \ref{Spectra}(b) shows the same for SCRO, at 50 and 660 K. The observed features for SCRO are significantly broader than for BFRO, probably due to the larger degree of antisite disorder. Still, the $A_{1g}$ symmetric stretching mode could be also clearly identified at $\sim 600$ cm$^{-1}$. A remarkable shift of this mode is observed on cooling for both studied compounds. In addition, a splitting of the $F_{2g}$ mode is observed at low temperatures for BFRO (see inset of Fig. \ref{Spectra}(a)). For SCRO, only a broad band is observed in the $\sim 400$ cm$^{-1}$ region, therefore the evolution of the $F_{2g}$ mode with temperature could not be reliably analyzed.

\begin{figure}
\includegraphics[width=0.9 \textwidth]{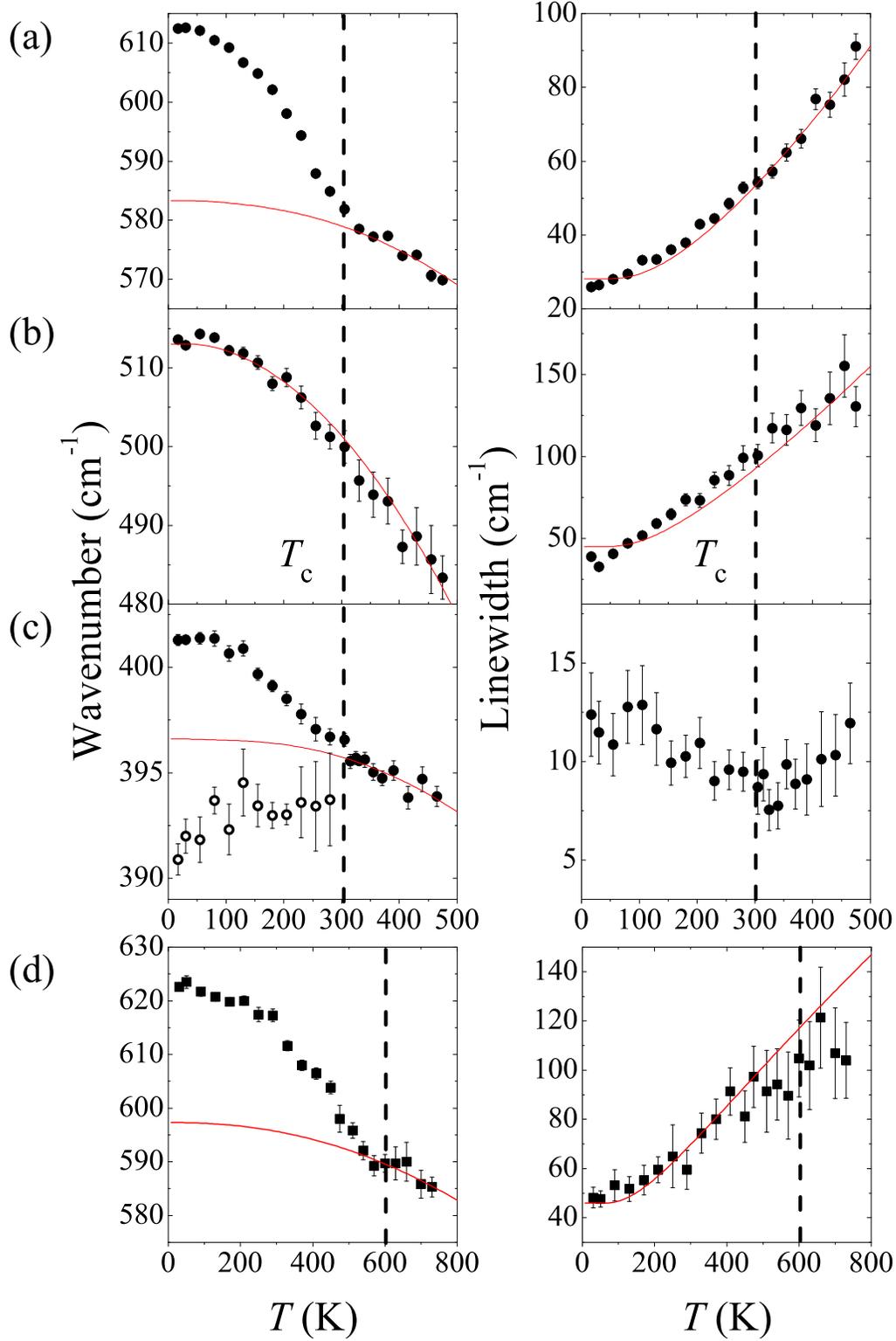}
\caption{\label{freq-Ba} (Color online) (a-c) Temperature dependence of the position and linewidth of the $A_{1g}$, $E_g$ and $F_{2g}$ modes of Ba$_2$FeReO$_6$. (d) Similar to (a), for the $A_{1g}$ mode for Sr$_2$CrReO$_6$. The solid curves give a fit of the conventional behavior above $T_c$, which is extrapolated below $T_c$ \cite{Balkanski}.}
\end{figure}

Figures \ref{freq-Ba}(a-c) show the $T$-dependent position and linewidth of the three observed Raman modes of BFRO, obtained from fits to Lorentzian profiles. Figure \ref{freq-Ba}(d) shows the same for the $A_{1g}$ mode of SCRO. The solid red curves represent the expected behavior for an anharmonic phonon decay through phonon-phonon scattering \cite{Balkanski}. For BFRO, the $A_{1g}$ mode shows a large anomalous hardening ($\sim 30$ cm$^{-1}$) below $T_c$, while the $E_g$ mode presents a conventional behavior. For SCRO, the $A_{1g}$ mode also shows an anomalous hardening of $\sim 25$ cm$^{-1}$ on cooling below $T_c$. For the $F_{2g}$ mode of BFRO, the analysis was performed under two distinct procedures. First of all, this mode was fitted with a single Lorentzian peak for all $T$. The right panel of Fig. \ref{freq-Ba}(c) shows the $T$-dependence of the obtained linewidth, with a clear broadening below $T_c=305$ K. As seen in the inset of Fig. \ref{Spectra}(a), this apparent broadening is actually a peak splitting that becomes unresolved close to $T_c$. Another analysis was then performed using two peaks for the fits below $T_c$. The left panel of Fig. \ref{freq-Ba}(c) shows the obtained peak positions. 

\begin{figure}
\includegraphics[width=0.9 \textwidth]{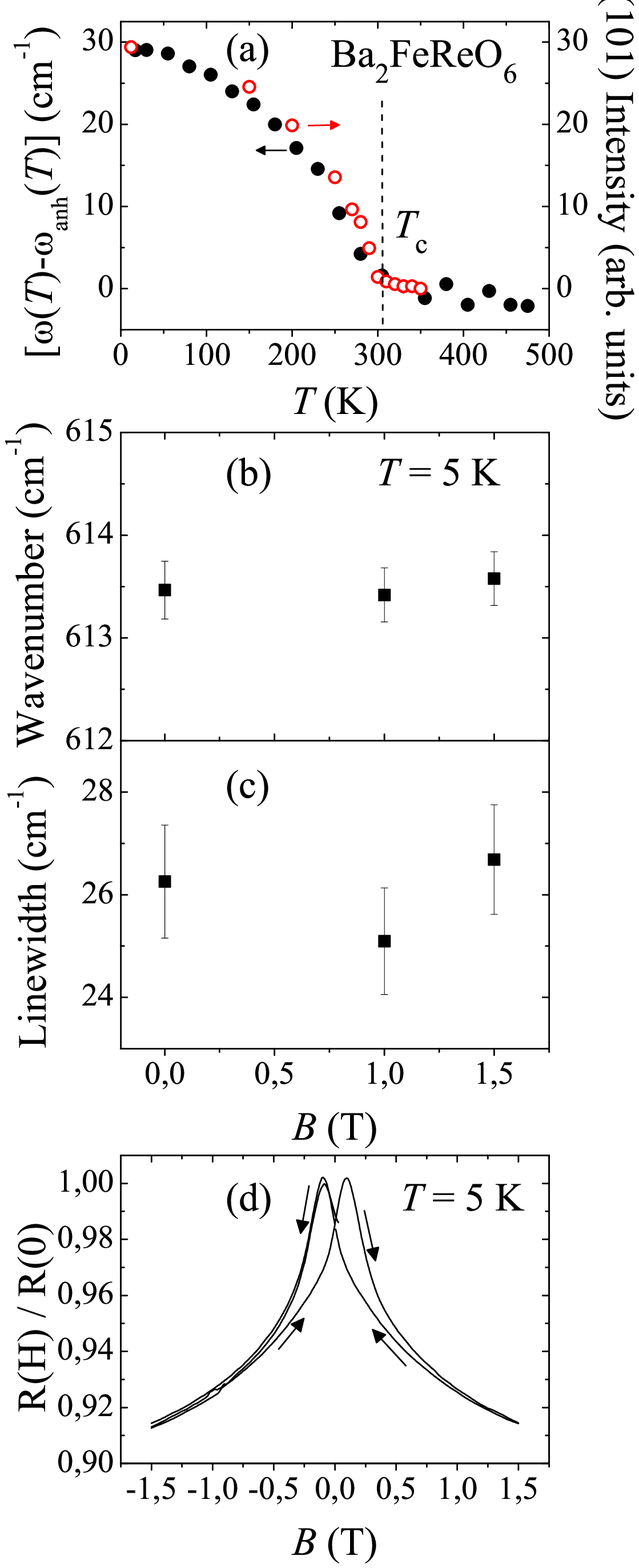}
\caption{\label{field} (Color online) (a) Comparison between the phonon renormalization [$\Delta\omega(T)$] and the neutron intensity of the (101) magnetic reflection of Ba$_{2}$FeReO$_{6}$. The dashed vertical line marks the magnetic ordering transition at $T_{C} \sim$305 K. Magnetic field dependence of (b) position and (c) linewidth of the $A_{1g}$ symmetric stretching mode and (d) normalized electric resistance of Ba$_{2}$FeReO$_{6}$ at 5 K.}
\end{figure}

The difference between the observed $A_{1g}$ phonon frequency and the conventional anharmonic behavior for BFRO is given in Fig. \ref{field}(a), as well as the neutron intensity of the (101) magnetic reflection that is proportional to [$\vec{\mu}$(Fe)$-\vec{\mu}$(Re)]$^2$ (Ref. \cite{Azimonte}). The scaling of both curves is remarkable, further supporting the magnetic origin of the $A_{1g}$ phonon anomaly. The magnetic field dependence of the $A_{1g}$ phonon position and linewidth of BFRO at 5 K were also investigated and are shown in Figs. \ref{field}(b) and \ref{field}(c), respectively, with no observable changes. Intergrain tunelling magnetoresistance measurements given in Fig. \ref{field}(d) indicate a reorientation of the magnetic domains parallel to applied fields consistent with previous results \cite{Prellier,Teresa1}. Since this experiment was performed in a pelletized polycrystalline sample, the magnetic moments must be reoriented without a physical rotation of the grains. We conclude that the $A_{1g}$ phonon energy is insensitive to the spin orientation with respect to the crystalline axes of each grain.
   
The phonon renormalization associated with magnetic ordering is normally interpreted in terms of the conventional spin-phonon coupling effect \cite{Baltensperger,Granado}. This coupling occurs if the exchange integrals are sensitive to a normal coordinate, leading to magnetic contributions to the harmonic energy of the lattice. One of the signatures of this effect is the proportionality between the anomalous phonon shifts and the square of the spontaneous sublattice magnetization \cite{Granado}, consistent with Fig. \ref{field}(a). 
For the case of oxygen stretching modes in the quasi-cubic double perovskites, each oxygen connects a specific pair of nearest-neighbor $M$-Re magnetic ions (see Fig. \ref{structure}). If one assumes that the spin-phonon coupling of the $A_{1g}$ mode is due to nearest-neighbor $J_{M-Re}$ modulation by the oxygen stretching vibration, each ($M$, Re) spin pair contributes to a term  $\delta H = - 1/2 (\partial ^2 J_{M-Re} / \partial x^2)x^2 \left\langle \vec{S}_{M}.\vec{S}_{Re}\right\rangle$ to the elastic energy, where $x$ is the coordinate of the oxygen ion along the line connecting $M$ and Re ions. Thus, each individual oxygen would contribute independently to the spin-phonon coupling, and both symmetric and anti-symmetric stretching modes should present identical energy anomalies. However, this is inconsistent with our observation for BFRO, in which the $E_g$ anti-symmetric stretching mode does not show any observable Raman shift anomaly at $T_c$ (see Figs. \ref{Spectra} and \ref{freq-Ba}(b)), in stark contrast to the $A_{1g}$ anomaly. Based on this consideration, we discard the modulation of $J_{Fe-Re}$ as the main source of the $A_{1g}$ energy renormalization for BFRO.

The structural transition at $T_s = T_c$ for BFRO associated with the strong spin-orbit coupling of Re $5d$ electrons \cite{Azimonte} might be also considered as a possible cause for the observed $A_{1g}$ phonon frequency anomaly. However, for SCRO the structural transition occurs at $T_s=260$ K $\neq T_c$ \cite{Teresa3}, while the $A_{1g}$ phonon energy anomaly occurs below $T_c \sim 605$ K, excluding this possibility. On the other hand, the splitting of the $F_{2g}$ triply degenerate bending mode in two peaks below $T_s$ for BFRO (see Fig. \ref{freq-Ba}(c)) is most likely related with the symmetry lowering to a tetragonal unit cell \cite{Azimonte}.

The coupling of the $A_{1g}$ mode with the Re and $M$ local moments actually occurs through the electron-phonon interaction. The symmetric stretching mode modulates the chemical pressure of the oxygen octahedra into the Re and $M$ ions. Since the Fermi electrons show mixed Re $5d$, O $2p$, and $M$ $3d$ character, the occupation of each of these orbitals is related to such chemical pressure. Indeed, when the oxygen cage is closer to the Re ions, the Re ($M$) $5d$ $(3d)$ occupation is expected to be lower (higher) with respect to the equilibrium oxygen positions, due to the Coulomb repulsion between the Fermi electrons and the O$^{2-}$ ions. Thus, the $A_{1g}$ mode is able to modulate the Re $5d$ and $M$ $3d$ occupations. Since these are half-metallic materials, the magnitudes of the local Re and $M$ spins will consequently be also modulated. Indeed, a smaller density of Re $5d$ electrons and a correspondly larger density of minority-spin $M$ $3d$ electrons, associated with an instantaneously compressed ReO$_6$ oxygen cage by the $A_{1g}$ mode, will reduce both Re and $M$ local spin magnitudes with respect to the equilibrium value. This effect does not occur for the antisymmetric stretching mode because this mode does not modulate the volume of the oxygen octahedra and the chemical pressure into the transition-metal ions (see Fig. \ref{structure}).

The record-high energy renormalization of the breathing lattice mode below $T_c$ can be rationalized using the coupling of this mode with the local moments argued above. While in the conventional formulation of the spin-phonon coupling \cite{Baltensperger,Granado} the phonon renormalization occurs solely due to the sensitivity of the exchange integrals to the normal coordinate $\eta$, in the present case also the spin magnitudes are functions of $\eta$ for the $A_{1g}$ mode. This leads to an additional pathway  to the phonon renormalization. The essential physics is captured by a simple ground state calculation for the magnetic exchange energy of the collinear ferrimagnetic state, using the magnetic energy per formula unit $E_{\eta}=-C_1 S_{M}^2 - C_2 S_{M}S_{Re}$, where $C_1=3(2J_1^{M-M}+J_2^{M-M})$ and $C_2 = 3J_1^{M-Re}$. The exchange integrals $J_1^{M-M}$, $J_2^{M-M}$ and $J_1^{M-Re}$ are represented in Fig. \ref{structure}.
The half-metallic states of BFRO and SCRO imply the condition $S_{Fe}(\eta)=S-S_{Re}(\eta)$, where $S$ is the constant net spin. Expanding $S_{Re}(\eta)$ in power series, the quadratic correction to the exchange energy at $T=0$ is:
\begin{equation}
\delta E^{(2)}(\eta)= \{ -C_1 [(S^{\prime 0} _{Re})^2 - (S-S^0 _{Re}) S^{\prime \prime 0} _{Re} ] - C_2 [-(S^{\prime 0} _{Re})^2 + (S/2-S^0 _{Re}) S^{\prime \prime 0} _{Re} ] \} \eta ^2, 
\end{equation}

\noindent where $S^{\prime 0} _{Re} \equiv dS_{Re}/d \eta (\eta=0)$ and $S^{\prime \prime 0} _{Re} \equiv d^2 S_{Re}/d \eta ^2 (\eta=0)$. This term adds to the lattice harmonic potential and renormalizes the phonon energy. For $T>T_c$, the exchange energy is null, while for ($0<T<T_c$) the orientational disorder caused by the thermally-activated spin waves must be taken into account, and the phonon renormalization is proportional to the spin correlation function such as in the conventional spin-phonon coupling \cite{Baltensperger,Granado}, consistent with our results (see Fig. \ref{field}(a)). The very large effect observed for the $A_{1g}$ mode is due to a strong sensitivity of the Re/$M$ local moments to the chemical pressure into the Re and $M$ sites, caused by the presence of itinerant spin-polarized electrons with mixed $M$ $3d$ and Re $5d$ character.

In summary, Raman scattering experiments in BFRO and SCRO show a large hardening of the $A_{1g}$ symmetric stretching mode of the oxygen octahedra below $T_c$. The remarkable magnitude of this effect and a comparison with the behavior of the anti-symmetric stretching mode shows this is not associated with a conventional spin-phonon coupling mechanism. A sensitivity of the local Re and Fe/Cr electronic densities and spin moments to the $A_{1g}$ lattice mode is inferred from our data, implying that a novel collective excitation involving the lattice, spin, and charge degrees of freedom is formed. This excitation is possibly not restricted to Re-based double perovskites, being rather present in other materials in which local electronic and spin densities are strongly dependent of a normal coordinate.

\begin{acknowledgments}

We thank D.O. Souza and J. Depianti for support in the synthesis of SCRO. LNLS is acknowledged for concession of beamtime. This work was supported by Fapesp, CNPq and CAPES, Brazil.

\end{acknowledgments}

\end{document}